\documentclass[aps,prl,twocolumn,superscriptaddress,floatfix]{revtex4}
\usepackage{graphicx}
\usepackage{dcolumn}
\usepackage{bm}
\usepackage{times}
\usepackage{color}
\usepackage{appendix}

\begin{document}
\bibliographystyle{apsrev}
\title{Domain wall pinning and hard magnetic phase in Co-doped bulk single crystalline Fe$_3$GeTe$_2$}

\author{Cong-Kuan Tian}
\author{Cong Wang}
\author{Wei Ji}
\author{Jin-Chen Wang}
\author{Tian-Long Xia}
\author{Le Wang}
\author{Juan-Juan Liu}
\author{Hong-Xia Zhang}
\author{Peng Cheng}\email{pcheng@ruc.edu.cn}

\affiliation{Department of Physics, Renmin University of China, Beijing 100872, P. R. China}
\affiliation{Beijing Key Laboratory of Opto-electronic Functional Materials $\&$ Micro-nano Devices, Renmin University of China, Beijing 100872, P. R. China}

\date{\today}

\begin{abstract}
We report the effects of cobalt doping on the magnetic properties
of two-dimensional van der Waals ferromagnet Fe$_3$GeTe$_2$.
Single crystals of (Fe$_{1-x}$Co$_x$)$_3$GeTe$_2$ with x=0-0.78
were successfully synthesized and characterized with x-ray
diffraction, energy dispersive x-ray spectroscopy and
magnetization measurements. Both the Curie-Weiss temperature and
ferromagnetic (FM) ordered moment of Fe$_3$GeTe$_2$ are gradually
suppressed upon Co doping. A kink in zero-field-cooling low field
M(T) curve which is previously explained as an antiferromagnetic
transition is observed for samples with x=0-0.58. Our detailed
magnetization measurements and theoretical calculations strongly
suggest that this kink is originated from the pinning of magnetic
domain walls. The domain pinning effects are suddenly enhanced
when the doping concentration of cobalt is around 50\%, both the
coercive field H$_c$ and the magnetic remanence to saturated
magnetization ratio M$_R$/M$_S$ are largely improved and a hard
magnetic phase emerges in bulk single crystal samples. The strong
doping dependent magnetic properties suggest more spintronic
applications of Fe$_3$GeTe$_2$.
\end{abstract}
\maketitle

\section{Introduction}
Two-dimensional (2D) van der Waals (vdW) ferromagnetic materials
have recently drawn great attentions for their potential 2D
magnetic, magnetoelectric and magneto-optic
applications\cite{2013VDW,2015Du,Wang2016Raman,Lee2016Ising,Kuo2016Exfoliation,Du2017One,Piquemal2016Magnetic,Xing2017Electric,2016Device,2017Device,YANG}.
For example, the layer-dependent intrinsic 2D ferromagnetism has
been demonstrated in two insulating vdW materials
Cr$_2$Ge$_2$Te$_6$\cite{Cr2Ge2Te6} and CrI$_3$\cite{CrI3}. The
following application of CrI$_3$ in making spintronic devices has
revealed surprisingly giant tunnelling magnetoresistance and the
possibility to push magnetic information storage to the atomically
thin limit\cite{Song2018Giant}. Comparing with insulators, vdW
magnetic metals are preferred for building spintronic
heterostructures as their metallic nature enabling the interplay
of both spin and charge degrees of freedom.

Fe$_3$GeTe$_2$ (FGT) serves as a rare metallic example of
itinerant ferromagnetic vdW materials\cite{2006FGT,2013FGT}. It
has a hexagonal crystal structure with the layered Fe$_3$Ge
substructure sandwiched by two layers of Te atoms and a van der
Waals gap in between. Early research finds ferromagnetic order
with Fe moments aligned along the c axis below Curie temperature
in bulk FGT (T$_C$$\approx$160~K-230~K)\cite{2016MAY}. Recent
reports show that itinerant ferromagnetism persists in FGT down to
the monolayer with an out-of-plane magnetocrystalline anisotropy
and tunable FM characteristics, making FGT a promising candidate
for spintronic applications\cite{2018Nature,2018Nano}. According
to current reports, the bulk FGT single crystal has a
ferromagnetic state with very small magnetic remanence to
saturated magnetization (M$_R$/M$_S$) ratio and coercivity at all
temperatures which limit its application in spintronic
architectures\cite{2016JAP,2016MAY}. The only way to obtain a hard
magnetic phase is making either nanoflakes or few layer
samples\cite{2018Nature,2018Nano,2018Hard}. On the other hand, Yi
$et~al.$ suggest an antiferromagntic (AFM) transition below 150~K
for FGT based on the low field magnetization data and theoretical
calculations\cite{2017AFM}. Therefore it is still controversial if
there is an AFM ground state at low temperature.

Chemical substitution is an effective way to tune the properties
and probe the underlying physics of magnetic materials. We noticed
that both Fe$_3$GeTe$_2$ and Ni$_3$GeTe$_2$ form the same crystal
structure while the intermediate element Cobalt failed to form a
'Co$_3$GeTe$_2$' phase according to current reports. This is
unusual because normally the properties of cobalt such as
Pauling's electronegativity and ionic radius lie in the middle
between iron and nickel. It would be interesting to see how the
magnetic properties of Fe$_3$GeTe$_2$ can be tuned by Co doping,
which may also provides insights about the controversial ground
state of FGT.

In this paper, we report the magnetic properties of
(Fe$_{1-x}$Co$_x$)$_3$GeTe$_2$ single crystals with $x$=0-0.78.
Our results suggest the previously reported suspicious AFM-like
transition in FGT is actually caused by the movement of magnetic
domain walls in a pinning state. The domain pinning effect can be
largely enhanced by Co-doping, which induces an intrinsic hard
magnetic phase (M$_R$/M$_S$$\sim$0.9) in contrast with the soft
magnetic phase in undoped FGT.

\section{Experimental details}

\begin{figure*}[htbp]
\centering
\includegraphics[width=0.9\textwidth]{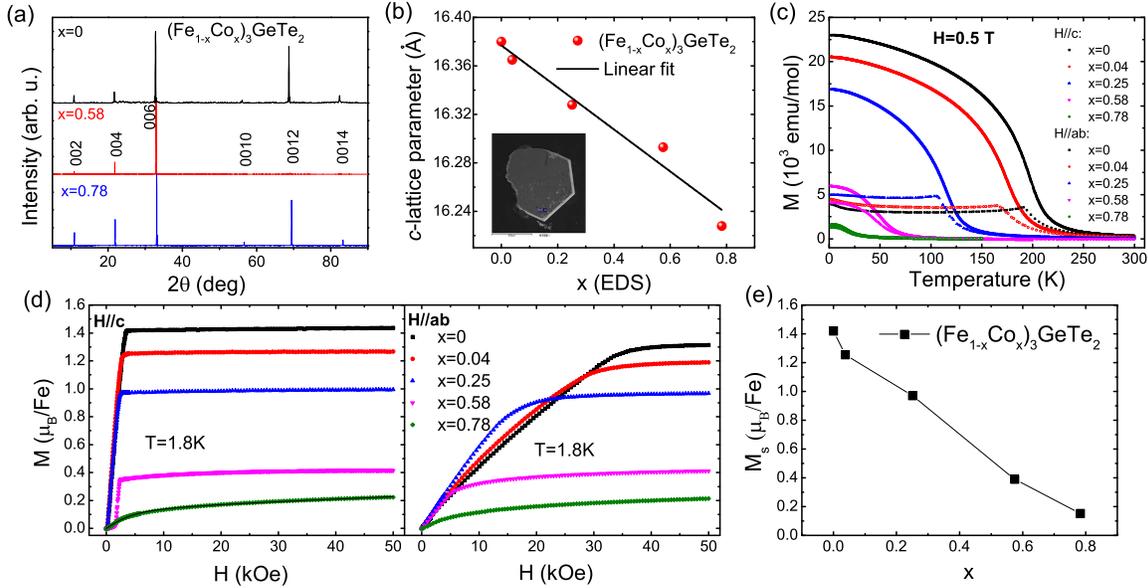}
\caption{(a) The x-ray diffraction patterns measured on single
crystals of (Fe$_{1-x}$Co$_x$)$_3$GeTe$_2$ showing (00L)
diffraction peaks. (b) Doping dependence of c-axis lattice
parameters. Inset shows one single crystal of x=0.58 imaging by a
scanning electron microscope. (c) Temperature dependence of the
magnetization M measured with H=0.5~T applied either parallel to
the $c$-axis (H$\parallel$c, solid symbols) or parallel to the
ab-plane (H$\parallel$ab, open symbols) for
(Fe$_{1-x}$Co$_x$)$_3$GeTe$_2$. (d) Isothermal magnetization
curves for different samples measured with H$\parallel$c and
H$\parallel$ab up to 5 T at T=1.8~K. The curve with x=0.78 is
fitted by equation (1). (e) Doping dependence of saturated
magnetic moment per Fe calculated from isothermal magnetization
curves.}
\end{figure*}

The single-crystalline samples of (Fe$_{1-x}$Co$_x$)$_3$GeTe$_2$
were prepared by the standard chemical vapor transport (CVT)
method with iodine as the transport agent similar to previous
reports\cite{2013FGT}. Crystals with typical dimensions of
1~mm$\times$1~mm$\times$0.1~mm are obtained with cobalt doping
values up to x=0.78. Further efforts in growing crystals with
larger $x$ failed and simply brought out products of CoTe$_{1.8}$
crystals. We characterized all samples with energy dispersive
x-ray spectroscopy (EDS, Oxford X-Max 50). The descriptions in
this paper about doping level $x$ all refer to the EDS values. The
single crystal x-ray diffraction patterns were collected from a
Bruker D8 Advance x-ray diffractometer using Cu K$_\alpha$
radiation. The magnetization measurements of our samples were
performed using a Quantum Design MPMS3.

We performed first-principles density functional theory
calculations using the same methods described in our previous
publications\cite{WC1,WC2,WC3}. In brief, we used a van der Waals
density functional (vdW-DF) method\cite{WC4,WC5}, with the optB86b
functional\cite{WC6} for the exchange part (optB86b-vdW) to
optimize atomic structures of bulk FGT, which usually reveals good
agreements of calculated structure-related properties with
experimental values of two-dimensional
materials\cite{WC7,WC8,WC9,WC10,WC11}. Given optimized structures,
we used the standard Perdew-Burke-Ernzerhof (PBE)
functional\cite{WC12} with the consideration of spin-orbit
coupling (SOC) to account energy differences of all considered
magnetic configurations, this scheme was found to share the
qualitatively same results with the Heyd-Scuseria-Ernzerhof
(HSE06) functional\cite{WC13,WC14} in other magnetic 2D layers,
e.g. CrI$_3$\cite{WC2} and CrSCl\cite{WC3}.

\section{Results and discussions}

Figure 1(a) presents the x-ray diffraction data of three single
crystals with $x$=0, $x$=0.58 and $x$=0.78 respectively. The peaks
can be indexed by (00L) with even values. No impurity peaks are
found within the instrument resolution. The $c$-axis lattice
parameters derived from the x-ray data decrease monotonically with
increasing $x$ as shown in Fig. 1(b). These results indicate the
successful introduction of cobalt into the FGT lattice.

Figure 1(c) shows the temperature dependent magnetization
measurements in zero-field-cooling (ZFC) with magnetic field of
0.5~T applied either parallel or perpendicular to the ab-plane.
The FM transition temperature T$_c$ of Fe$_3$GeTe$_2$ sample is
around 200~K, then it is gradually suppressed with Co-doping. On
the other hand the magnetic easy axis is along H$\parallel$c for
all samples while the magnetic anisotropy and the ordered moment
of Fe gradually decrease with increasing $x$.

The isothermal magnetization curves at T=1.8~K are presented in
Fig. 1(d). For crystals with x=0-0.58, the rapid saturated
magnetizations confirm their ferromagnetic ground states. For
x=0.78, the shape of M(H) curve resembles those observed in
cluster glasses\cite{Glassfit,2018Ni}. Therefore we fit the M(H)
curve with a modified Langevin function represented by
\begin{equation}\label{equ1}
M(H)=M_s L (\mu H /k_B T) + \chi H
\end{equation}
Here $\mu$ is the average moment per cluster, L(x)=coth(x)-1/x is
the Langevin function, M$_S$ is the saturated moment, and $\chi$
is the paramagnetic susceptibility\cite{Glassfit,2018Ni}. The
fitting result gives an M$_S$ value of 0.153~$\mu_B$ for x=0.78.
For other samples, the M$_S$ values were determined from the
intercept of a linear fit of H$>$1~T data with H=0. The doping
dependence of saturated magnetic moment per Fe/Co is shown in Fig.
1(e). The suppression of saturated moment is quite similar as that
in nickel-substituted Fe$_3$GeTe$_2$\cite{2018Ni}. One difference
is that the cluster glass behavior starts at x=0.37 for Ni doping
while the ferromagnetic state still seems to be robust at least
for x=0.58 in the case of Co doping.

\begin{figure}[htbp]
\centering
\includegraphics[width=0.48\textwidth]{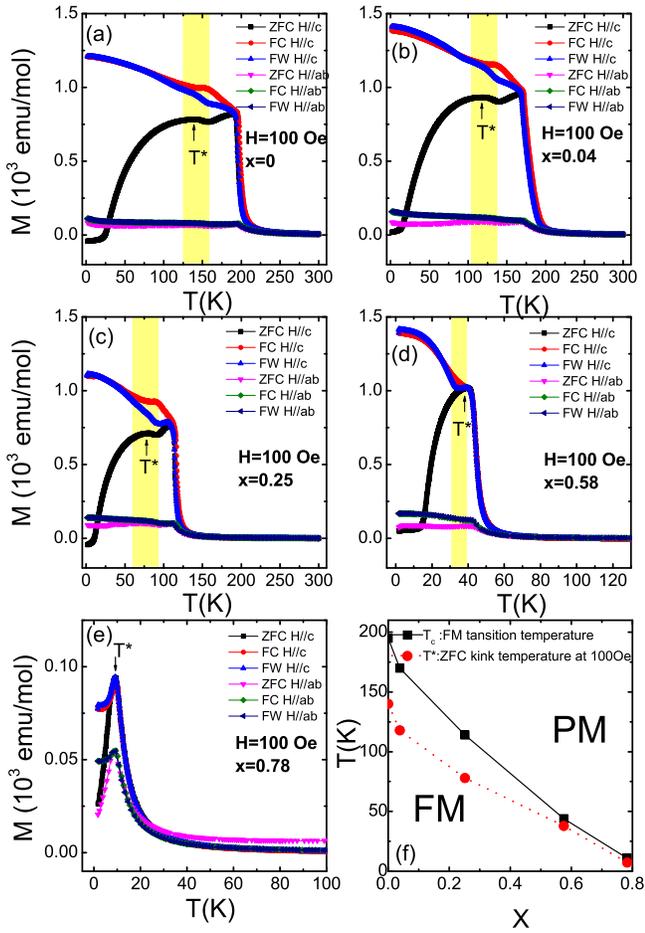}
\caption{(a)-(e) Temperature dependence of Magnetization for
x=0-0.78 with 100~Oe magnetic field applied in directions of
H$\parallel$c and H$\parallel$ab. (f) Temperature vs doping phase
diagram for (Fe$_{1-x}$Co$_x$)$_3$GeTe$_2$. The FM transition
temperature T$_c$ is defined by the minimum of dM/dT curve. The
T$^*$ is defined as the ZFC kink temperature at H=100~Oe, which
indicates a crossover from freely moved magnetic domains to pinned
ones under this magnetic field.}
\end{figure}

When the temperature dependent magnetizations are measured at a
lower magnetic field such as 100~Oe, anomalous AFM-like kinks
emerge in the ZFC M(T) curves with H$\parallel$c as shown in Fig.
2(a)-(e). For Fe$_3$GeTe$_2$, the kink temperature T$^*$ is around
150~K and the ZFC magnetizations approaches zero below 30~K which
is lower than the counterpart in the H$\parallel$ab ZFC curve.
Meanwhile a thermo-hysteresis is observed for the field-cooling
(FC) and field-warming (FW) curves at around the kink temperature.
The similar phenomenon has been reported previously and explained
as a new AFM transition at the kink temperature (antiparallel spin
arrangement along the $c$-axis between different Fe$_3$Ge
layers)\cite{2017AFM}. Another report explained this phenomenon as
a Kondo scenario coherent-incoherent crossover which is related to
the hybridization between local moments and conduction
electrons\cite{2018Kondo}. We find that this crossover or
transition remains in Co-doped samples up to x=0.58 with occurring
temperature T$^*$ approaching the FM transition temperature. For
x=0.78 all M(T) curves show peaks at T=9~K, which is possibly due
to the formation of cluster spin glass. In Fig. 2(f) the FM
transition temperature T$_c$(minimum in the dM/dT curve) and the
anomalous ZFC kink temperature T$^*$ at H=100~Oe are plotted as a
function of doping $x$.

\begin{figure}[htbp]
\centering
\includegraphics[width=0.48\textwidth]{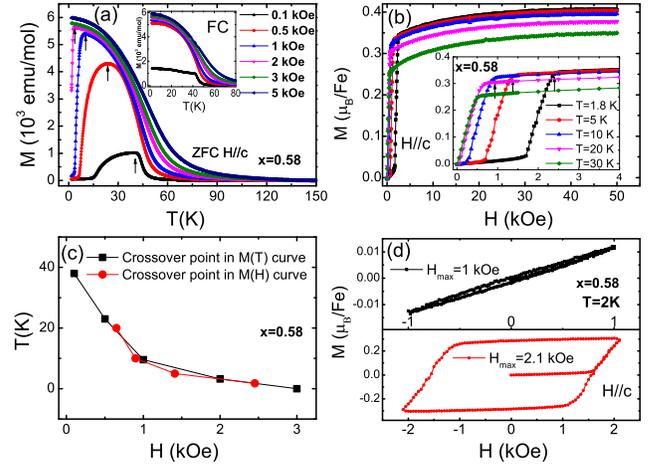}
\caption{(a) ZFC magnetization curves for x=0.58 under different
magnetic fields. Inset shows the corresponding FC magnetization
curves. (b) Isothermal magnetization curves for x=0.58 with
H$\parallel$c at different temperatures. Inset shows the enlarged
view of the low field data. (c) Crossover points in ZFC M(T)
curves and M(H) curves as marked by black arrows in (a) and (b)
can be scaled together in a temperature vs field plot. (d)
Hysteresis loops for x=0.58 with different maximum magnetic fields
applied parallel to c-axis at T=2~K.}
\end{figure}

In order to clarify the origin of T$^*$, the x=0.58 sample is
chosen for detailed magnetization measurements. Three major
features are found: (1) The kink gradually moves to low
temperature with increasing magnetic field and finally disappears
at H=3~kOe (Fig. 3(a)). (2) No kink is observed in FC curves under
the same field (inset of Fig. 3(a)). (3) The M(H) curve at T=1.8~K
with H$\parallel$c undergoes a steep magnetization jump at
H$\approx$2~kOe as shown in the inset of Fig. 3(b). This jump
gradually moves to lower field and finally disappears at T=30~K.
The kinks in M(T) curve and the jumps in M(H) curve (marked by
black arrows in Fig. 3(a) and Fig. 3(b) respectively) can actually
be scaled together if we plot their occurring temperature and
field in Fig. 3(c), indicating they should have the same origin.

Based on the above observations, there are two possible
explanations for the kinks and jumps mentioned above, namely a
spin-flop transition (from AFM to FM) or a pinning-depinning
crossover of magnetic domain walls. We argue that a spin-flop
transition is unlikely for two reasons. First of all, according to
our theoretical calculations described in the previous section,
the interlayer FM configuration is 0.81 meV/Fe more stable than
the interlayer AFM configuration, suggesting a FM
groundstate,which indicates flipping of magnetic moment from an
anti-parallel to a parallel configuration is, most likely, not a
reason for the observed magnetic transition. Even if the
interlayer magnetism appears to be AFM, owing some reason, e.g. a
particular stacking\cite{WC2}, the 0.81 meV energy difference
implies that it may take roughly 10 Tesla to flip the interlayer
magnetic moment, roughly two orders of magnitude larger than the 2
kOe field we observed in our experiment. Magnetic field at this
strength would more likely to cause a movement or depinning of
magnetic domains, rather than flop the spins. Secondly, the
magnetization loop with maximum field of 1~kOe exhibits a linear
feature with very weak hysteresis, while significant FM hysteresis
appears in the loop with maximum field of 2.1~kOe (Fig. 3(d)).
This means that if a spin-flop transition from AFM to FM really
exists, it can not be tuned back when the field is cooled from
2.1~kOe. This behavior clearly contradicts the common features of
spin-flop transitions.

So we propose a crossover from pinning to depinning of magnetic
domain walls as the reason for the magnetization kinks shown in
Fig. 2 and Fig. 3. When the sample is cooled under zero field, the
magnetic domains start to be pinned below the crossover
temperature T$^*$ with their total moment close to zero (keeping
the lowest magnetostatic energy). Then applying a low field of
100~Oe at lowest temperature is not enough to move the pinned
domains. With increasing temperature, thermal fluctuations
gradually weaken the pinning force and finally completely depin
the domains above T$^*$ with domain moment well aligned along the
field direction. This explains why the kink of magnetization with
deceasing temperature never occurs in FC curve. Because in FC
process the domains are always pinned with the effective FM moment
aligned along the cooling field. The thermo-hysteresis observed in
the FC and FW curves is likely due to the domain structure
dynamics when switching between pinning and depinning state. A
recent scanning tunneling microscope (STM) study on FGT uses
ferromagnetic Ni tips to mimic the FC and FW
process\cite{2018STM}. The data show that the domain structure in
FC process is different from that in FW process even at the same
temperature, which naturally explains the
thermal-hysteresis\cite{2018STM}. It should be mentioned that a
spin-flop transition could also possibly generate the
thermo-hysteresis\cite{2017AFM}, however previous neutron
scattering studies on FGT do not support an AFM spin-configuration
at low temperature\cite{2016MAY}.

The hysteresis loops are measured for all samples and reveal new
doping induced magnetic properties. As shown in Fig. 4(a) and
inset, it is evident that all (Fe$_{1-x}$Co$_x$)$_3$GeTe$_2$
samples with x$\leq$0.58 are pinning type ferromagnets. Namely the
initial magnetization of the sample is negligible but suddenly
become significant beyond a certain field, this change in
magnetization is reached by the movement of the pinned domain
walls\cite{2016JAP}. For x=0-0.25, both the coercive field H$_c$
($\sim$200 Oe) and the magnetic remanence to saturated
magnetization ratio M$_R$/M$_S$ ($<$0.1) are very low, which
belong to soft magnetic properties same as previous reports about
Fe$_3$GeTe$_2$\cite{2016JAP,2016MAY}. However for samples with
0.46$\leq$x$\leq$0.58, the hysteresis loops suddenly display a
near square shape with greatly enhanced coercivities (Fig. 4(b),
coercive field$\sim$1.5~kOe). Meanwhile the calculated M$_R$/M$_S$
ratios are all larger than 0.8 from x=0.46 to x=0.58 with maximum
value of 0.9 (Fig. 4(c)). These are all well-defined hard magnetic
properties similar as that in the previously reported few layer
samples or thin films of FGT\cite{2018Nature,2018Nano,2018Hard}.
These results suggest hard magnetic phases can also be induced by
Co doping in bulk single crystals. For samples with x$\geq$0.68,
both the hard magnetic properties and pinning type magnet features
gradually disappear. To summarize the results in Fig. 4, we have
discovered that the coercive fields and M$_R$/M$_S$ values in
(Fe$_{1-x}$Co$_x$)$_3$GeTe$_2$ are strongly doping dependent, hard
magnetic phases can be realized at 0.46$\leq$x$\leq$0.58.

\begin{figure}[htbp]
\centering
\includegraphics[width=0.48\textwidth]{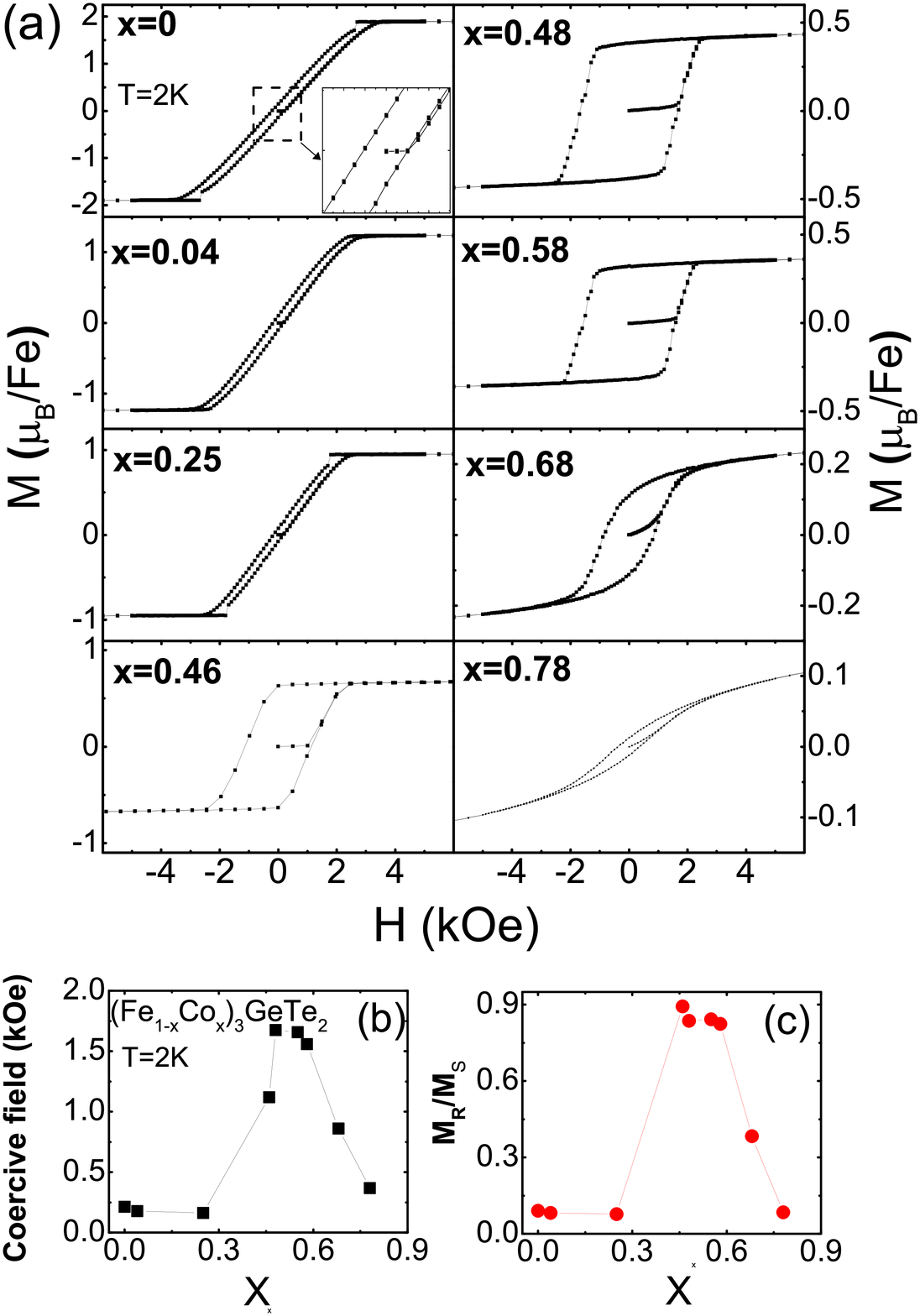}
\caption{(a) Hysteresis loops for samples with different doping
concentrations at T=2~K and H$\parallel$c. The inset shows the
enlarged view of the dashed box area. The data were measured with
H$_{max}$=$\pm$50~kOe. Doping dependent of coercive fields (b) and
M$_R$/M$_S$ values (c) at T=2~K for
(Fe$_{1-x}$Co$_x$)$_3$GeTe$_2$. The data are calculated from
hysteresis loops.}
\end{figure}

A major source of hysteresis in ferromagnets is the pinning of
magnetic domain walls\cite{2018Pin}. Generally speaking in order
to get a high coercive field H$_c$ in a pinning type magnet, it
requires the formation of a large domain wall energy (DWE) and an
effective network of pinning centers capable of locally increasing
DWE to inhibit the domain wall movement\cite{2016JAP}. The doping
of Co should somehow greatly improve the DWE of Fe$_3$GeTe$_2$
thus induces hard magnetic properties. It should be mentioned that
this improvement of DWE seems to only occur when
Fe:Co$\approx$1:1. Samples with x$\leq$0.25 and x$\geq$0.68 all
exhibit soft magnetic properties. We have repeated the above
findings on more samples with slightly different synthesize
procedures and nominal doping, the results show that the emergence
of hard magnetic phase only depends on the doping concentration.
Hard magnetic properties are crucial for the applications of 2D
magnetic materials in spintronics. We have shown the possibility
of getting a tunable hard magnetic phase through chemical doping
in Fe$_3$GeTe$_2$ bulk single crystals instead of making few layer
samples or thin films. These findings should shed new light on the
research and application of itinerant 2D vdW ferromagnetic metal
Fe$_3$GeTe$_2$.

\section{Conclusions}

In summary, a series of (Fe$_{1-x}$Co$_x$)$_3$GeTe$_2$ (x=0-0.78)
single crystals have been successfully grown by CVT methods. All
samples with x=0-0.58 are pinning type magnets and the previously
reported AFM-like transition in Fe$_3$GeTe$_2$ should originate
from the movement of pinned magnetic domain walls based on our
data analysis. The coercive fields and M$_R$/M$_S$ values are
strongly doping dependent. Instead of making few layer samples,
the hard magnetic properties can be realized in bulk single
crystals of Fe$_3$GeTe$_2$ with Co doping.

\section{Acknowledgments}
The authors thank the helpful discussion with Prof. Yu Ye, Prof.
Wei Bao and Prof. Lei Shan. This work is supported by the National
Natural Science Foundation of China (No. 11227906 and No.
11204373).

\bibliography{Bibtex}
\end{document}